\documentclass[12pt,preprint]{aastex}

\newcommand{\etal  }{{et al.} }
\newcommand{\msun}{\thinspace M_\odot}  
\newcommand{\rsun}{\thinspace R_\odot}  
\newcommand{\vect}[1]{\mbox{\boldmath$#1$}}

\def\lesssim{\mathrel{\hbox{\rlap{\hbox{\lower4pt\hbox{$\sim$}}}\hbox{$<$}}}}
\def\gtrsim{\mathrel{\hbox{\rlap{\hbox{\lower4pt\hbox{$\sim$}}}\hbox{$>$}}}}
\newcommand{\cm  }{\,{\rm cm}^{-3} }

\newcommand{\dfrac}[2]{{\displaystyle \frac{#1}{#2}}  }

\shorttitle{Second Core Formation and High Speed Jets}
\shortauthors{Machida  \etal 2006}

\begin{document}

\title{Second Core Formation and High Speed Jets: Resistive MHD Nested Grid Simulations}

\author{Masahiro N. Machida\altaffilmark{1} and Shu-ichiro Inutsuka\altaffilmark{2}} 
\affil{Department of Physics, Graduate School of Science, Kyoto University, Sakyo-ku, Kyoto 606-8502, Japan}
\and
\author{Tomoaki Matsumoto\altaffilmark{3}}
\affil{Faculty of Humanity and Environment, Hosei University, Fujimi, Chiyoda-ku, Tokyo 102-8160, Japan}

\altaffiltext{1}{machidam@scphys.kyoto-u.ac.jp}
\altaffiltext{2}{inutsuka@tap.scphys.kyoto-u.ac.jp}
\altaffiltext{3}{matsu@i.hosei.ac.jp}

\begin{abstract}
The stellar core formation and high speed jets driven by the formed
 core are studied by using three-dimensional resistive
 magnetohydrodynamical (MHD) nested grid simulations.
Starting with a Bonnor-Ebert isothermal cloud rotating in a uniform
 magnetic field, we calculate the cloud evolution from the molecular
 cloud core ($n_c = 10^6\cm$, $r_c = 4.6 \times 10^4$ AU) to the
 stellar core ($n_c \simeq 10^{23} \cm$, $r_c \simeq 1 \rsun$),
 where $n_c$ and $r_c$ denote the central density and radius of the
 objects, respectively.
We resolve cloud structure over 7 orders of magnitude in spatial
 extent and over 17 orders of magnitude in density contrast.
For comparison, we calculate two models:
 resistive and ideal MHD models.
Both models have the same initial condition,
 but the former includes dissipation process of magnetic field
 while the latter does not.
The magnetic fluxes in  resistive MHD model are extracted from
 the first core during $10^{12} \cm < n_c < 10^{16} \cm$ by
 Ohmic dissipation.
Magnetic flux density of the formed stellar core
 ($n_c \simeq 10^{20} \cm$) in  resistive MHD model is two orders of
 magnitude smaller than that in ideal MHD model.
Since magnetic braking is less effective in  resistive MHD model,
 rapidly rotating stellar core (the second core) is  formed.
After stellar core formation, the magnetic field of the core is
 largely amplified both by magneto-rotational instability and
 the shearing motion between the stellar core and ambient medium.
As a consequence,
 high speed ($\simeq 45$ km\,s$^{-1}$) jets are
 driven by the second core,
 which results in strong mass ejection.
A cocoon-like structure around the second core
 also forms with clear bow shocks.

\end{abstract}


\section{Introduction}
Outflows and jets are ubiquitous in star-forming regions and closely related to the accretion process, angular momentum transportation and magnetic field in the cloud.
Recent observations indicate these flows are composed of multiple components in space and velocity \citep{pyo03,pyo05}.
Molecular outflows observed by CO emission have speed of $\sim 10$ km\,s$^{-1}$  and wide opening angle \citep[e.g.,][]{belloche02}, while jets observed by SiO and FeII emissions have speed of $100$ km\,s$^{-1}$ and  well collimated structure \citep[e.g.,][]{hirano06}.
These observations suggest that different flows which have different sizes, flow speeds, and opening angles are driven by similar mechanism but with different conditions, or at different evolutional phases.
Although internal structure and rotational motion inside the outflows and jets are observed by recent high-resolution telescopes \citep{bacciotti02,choi05}, driven points of these flows are not observed because these regions are embedded in the dense cloud core and located in the very vicinity of the protostar where we can not spatially resolve.
Thus, the driving mechanism of outflows and jets are not understood well.
Since the effect of radiation pressure is small in low mass stars,  these flows are supposed to be driven by the Lorentz force.
The magnetic energies observed by Zeeman effect measurements are comparable or larger than the thermal energy \citep{crutcher99},  implying that magnetic effects are important for the evolution of molecular clouds. 
Thus, numerical MHD simulations are needed in order to understand the formation processes of protostars, jets and outflows.

Assuming spherical symmetry, the evolutions from the molecular cloud  to the stellar core are calculated by many authors \citep[e.g.,][]{larson69,winkler80a,winkler80b,masunaga98,masunaga00}, and they found density, velocity and temperature structures of the cloud and stellar core.
\citet{bate98} and \citet{whitehouse06} have calculated stellar core formation from the molecular cloud  core in their three dimensional SPH simulations.
In their calculations, the spiral or bar structures are appeared at high density, because the bar mode instability is induced in rapidly rotating cores  \citep{durisen86}.
 Since angular momentum is effectively removed due to non-axisymmetric structure, the central region contracts and stellar core forms. 
 However, angular momentum is largely removed by the magnetic braking \citep{basu94,machida05a} and by driving outflows \citep{matsu04,machida04,machida05b}. 
 Therefore, they overestimate the angular momentum of the formed core, because they ignore the magnetic field in the cloud.
 The evolution of magnetized clouds are investigated by \citet{tomisaka98,tomisaka00,tomisaka02} in his two dimensional nested grid simulation.
 In his calculation, low-velocity (outflow; ${\rm v} \simeq2$km\,s$^{-1}$) and high-velocity (jet; ${\rm v} \simeq 30$ km\,s$^{-1}$) flows are driven by first core and second core, respectively. 
  However, he adopt ideal MHD approximation, which is valid in low-density gas region ($n \lesssim 10^{12}\cm$), however, not valid in high-density gas region ($n \gtrsim 10^{12}\cm$).
  \citet{nakano02} found significant magnetic flux loss occurs during $10^{12} \cm \lesssim n \lesssim 10^{15}\cm$ by Ohmic dissipation.
Therefore,  \citet{tomisaka98,tomisaka00,tomisaka02} overestimate the magnetic flux of the cloud especially in high-density gas region.

 In this paper, we calculate cloud evolution from cloud core ($n_c = 10^6\cm$, $r_c = 4.6 \times 10^4$ AU) to stellar core phase ($n_c \simeq 10^{23} \cm$, $r_c \simeq 1 \rsun$) using three dimensional resistive MHD nested grid method and study the formation process of high-speed jets.
 We also calculate cloud evolution using ideal MHD approximation, and compare the resistive MHD model with the ideal MHD model.

\section{Model and Numerical Method}
 Our initial settings are almost the same as those of \citet{machida06}.
 To study the cloud evolution, we use the three-dimensional resistive MHD nested grid code.
 We solve the resistive MHD equations including the self-gravity:  
\begin{eqnarray} 
& \dfrac{\partial \rho}{\partial t}  + \nabla \cdot (\rho \vect{v}) = 0, & \\
& \rho \dfrac{\partial \vect{v}}{\partial t} 
    + \rho(\vect{v} \cdot \nabla)\vect{v} =
    - \nabla P - \dfrac{1}{4 \pi} \vect{B} \times (\nabla \times \vect{B})
    - \rho \nabla \phi, & \\ 
& \dfrac{\partial \vect{B}}{\partial t} = 
   \nabla \times (\vect{v} \times \vect{B}) + \eta \nabla^2 \vect{B}, & \\
& \nabla^2 \phi = 4 \pi G \rho, &
\end{eqnarray}
 where $\rho$, $\vect{v}$, $P$, $\vect{B} $, $\eta$ and $\phi$ denote the density, 
velocity, pressure, magnetic flux density, resistivity and gravitational potential, respectively. 
 To mimic the temperature evolution calculated by \citet{masunaga00}, we adopt the piece-wise polytropic equation of state.
\begin{equation} 
P = \left\{
\begin{array}{ll}
 c_s^2 \rho & \rho < \rho_c, \\
 c_s^2 \rho_c \left( \frac{\rho}{\rho_c}\right)^{7/5} &\rho_c < \rho < \rho_d, \\
 c_s^2 \rho_c \left( \frac{\rho_d}{\rho_c}\right)^{7/5} \left( \frac{\rho}{\rho_d} \right)^{1.1}
 & \rho_d < \rho < \rho_e, \\
 c_s^2 \rho_c \left( \frac{\rho_d}{\rho_c}\right)^{7/5} \left( \frac{\rho_e}{\rho_d} \right)^{1.1}
 \left( \frac{\rho}{\rho_e}   \right)^{5/3}
 & \rho > \rho_e, 
\label{eq:eos}
\end{array}
\right.  
\end{equation}
 where $c_s = 190$\,m\,s$^{-1}$, 
$ \rho_c = 1.92 \times 10^{-13} \, \rm{g} \, \cm$ ($n_c = 5 \times 10^{10} \cm$), 
$ \rho_d = 3.84 \times 10^{-8} \, \rm{g} \, \cm$  ($n_d =  10^{16} \cm$), and
$ \rho_e = 1.92 \times 10^{-3} \, \rm{g} \, \cm$  ($n_e = 5 \times 10^{20} \cm$).
 In this paper, a spherical cloud with critical Bonnor-Ebert \citep{ebert55, bonnor56} density profile having $\rho_{c,0} = 3.84 \times 10^{-18} \, \rm{g} \, \cm$ ($n_{c,0} = 10^6\cm$) of the central (number) density is used  as the initial condition.
 Initially the cloud rotates rigidly ($\Omega_0 = 1.02 \times 10^{-13}$s$^{-1}$) around the $z$-axis and has uniform magnetic field ($B_0 =212 $$\mu$G) parallel to the $z$-axis (or rotation axis).
 To promote the contraction, we increase the density as by 30 \% from the critical Bonnor-Ebert sphere.

  We adopt the nested grid method  \citep[for detail, see][]{machida05a} to obtain high spatial resolution near the center.
  Each level of rectangular grid has the same number of cells ($ = 64 \times 64 \times 32 $),  but cell width $h(l)$ depends on the grid level $l$.
 The cell width is reduced to 1/2 with increasing  grid level ($l \rightarrow l+1$).
 The highest level of grids changes dynamically.
 We start the calculation with 4 grid levels ($l=1,2,3,4$).
 Box size of the initial finest grid $l=4$ is chosen equal to $2 R_{\rm c}$, where $R_c$  denotes the radius of the critical Bonnor-Ebert sphere. 
 The coarsest grid ($l=1$), therefore, has a box size equal to $2^4\, R_{\rm c}$. 
 A boundary condition is imposed at $r=2^4\, R_{\rm c}$, where the magnetic field and ambient gas rotate at an angular velocity of $\Omega_0$ (for detail see \citealt{matsu04}).
  A new finer grid is generated whenever the minimum local Jeans length 
$ \lambda _{\rm J} $ becomes smaller than $ 8\, h (l_{\rm max}) $, where $h$ is the cell width. The maximum level of grids is restricted to $l_{\rm max} = 21$.
 Since the density is highest in the finest grid, generation of new grid ensures the Jeans condition of \citet{truelove97} with a margin of a safety factor of 2.

 We calculate two models: (a) resistive and (b) ideal MHD models.
Both models have the same initial condition shown above.
The former include resistive term ($\eta \nabla^2 \vect{B}$ ) in induction equation (eq. [3] ), while the latter does not.

\section{Results}

\begin{figure}
\plotone{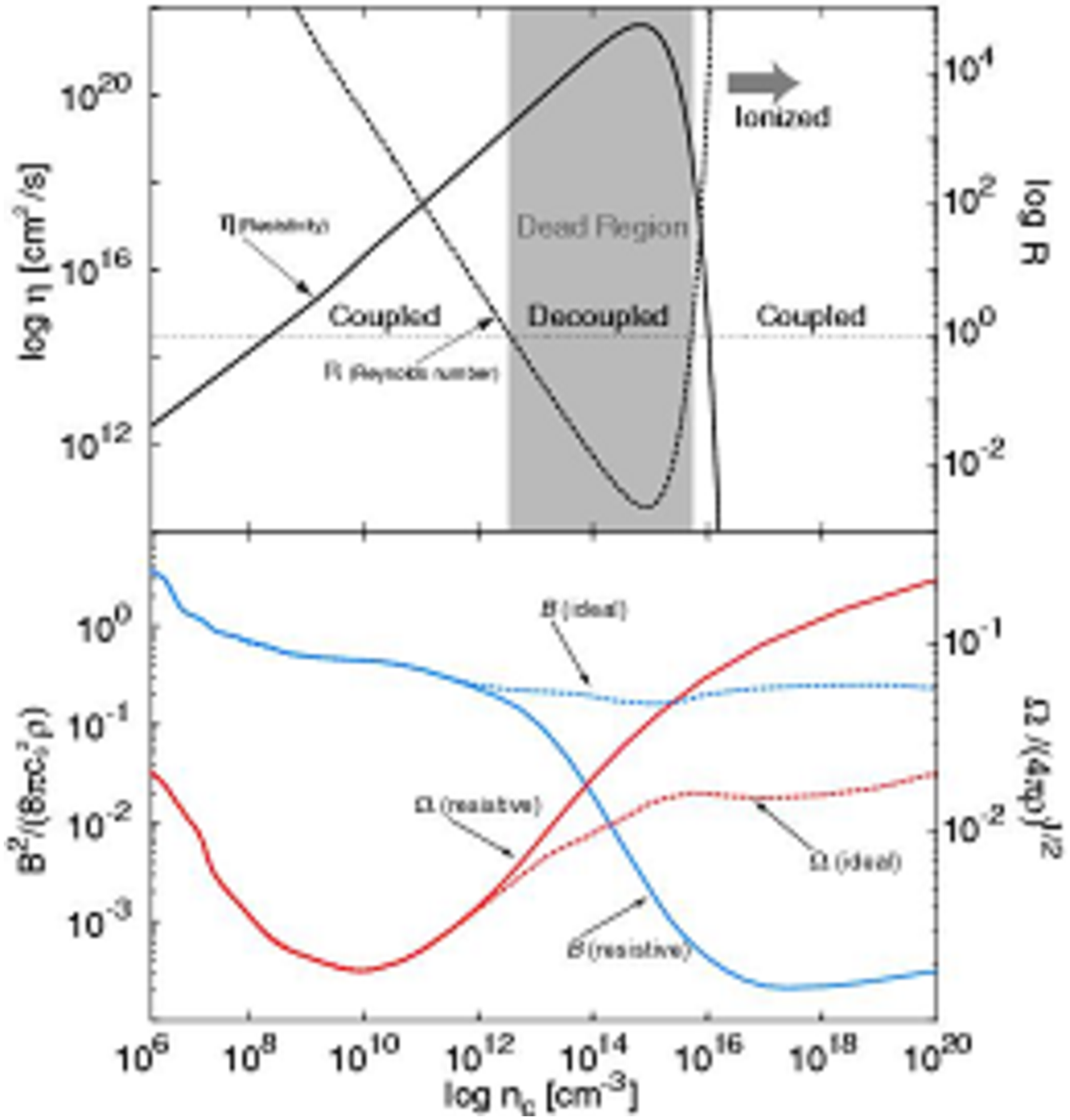}
\caption{
Upper panel: Resistivity ($\eta$; left axis) and magnetic reynolds number (R; right axis) as a function of the central number density.
The magnetic field well couple with the gas in ``coupled'' region, while magnetic field  decuples from the gas in ``decoupled'' region.
Lower panel:  The magnetic ($B^2/8\pi$) to thermal energy ($c_s^2\rho$) ratio (left axis ) within $n < 0.1\,n_{\rm c}$ and angular velocity ($\Omega$) normalized by free-fall timescale ([$4\pi G \rho_c]^{1/2}$) at the center (right axis) against central number density.
The gas begins to ionized when number density reaches $n\simeq 10^{15} \cm$.
}
\label{fig:1}
\end{figure}

 The molecular gas obeys the isothermal equation of state with temperature of $\sim 10$ K until $n_c \simeq 5 \times 10^{10}\cm$ ( isothermal phase), then cloud collapses almost adiabatically  ($5\times 10^{10} \lesssim n_c \lesssim 10^{16}$; adiabatic phase) and quasi-static core (i.e., first core)  forms during the adiabatic phase \citep{larson69, masunaga98}.
 In our calculations,  the first core forms both in resistive and ideal MHD models when the central density reaches $n_c \simeq 8\times10^{12} \cm$ and these cores have radius of $\simeq 10.6$ AU.
 The magnetic flux is expected to be removed from the  first core during the adiabatic phase \citep{nakano02}.
 We estimate the resistivity ($\eta$) according to   \citet{nakano02} and represented by a thick line in upper panel of Figure~\ref{fig:1} which has a peak at $n \simeq 10^{15} \cm$.
  We also estimate the magnetic Reynolds number ($R \equiv v_f \, \lambda_j \, \eta^{-1}$; dotted line in upper panel of Fig.~\ref{fig:1}) using the free-fall velocity ($v_f$) and Jeans length ($\lambda_J$).
 The dotted line indicate that magnetic dissipation is effective during $10^{12} \cm \lesssim n_c \lesssim 10^{16} \cm$  which corresponds to the results of \citet{nakano02}.

 The ratio of magnetic to thermal energy  within $n < 0.1\,n_{\rm c}$ and the angular velocity  normalized by the free-fall timescale  at the center are plotted against the central density in lower panel of Figure~\ref{fig:1}.
 The magnetic energy and angular velocity are almost the same during $10^6 \cm <n_c \lesssim 10^{12} \cm$ between resistive and ideal MHD models.
  The magnetic energy decreases in early isothermal phase ($10^6\cm < n_c \lesssim 10^8 \cm$), because the cloud collapses along magnetic field line \citep{machida05a}, then the magnetic energy is comparable to the thermal energy during $10^8\cm \lesssim n_c \lesssim 10^{12} \cm$.
 The ratio of magnetic to thermal energy in ideal MHD model keeps a constant value during $10^8 \cm \lesssim n_c \lesssim 10^{20}\cm$ and the magnetic field have equivalent energy to the thermal energy until final stage (the ratio of the magnetic to thermal energy is $\simeq 0.25$).
 The angular velocities both in resistive and ideal MHD models decrease during isothermal phase, because cloud collapses vertically and magnetic braking is effective in this phase as shown in \citet{machida05a}.
 The angular velocity in ideal MHD model begins to increase at $n\simeq 10^{10} \cm$, then it keeps an almost constant value ($\omega [4\pi G \rho]^{-1/2}\simeq 0.02$) until the final stage, while it continues to increase after the central density reaches  $n_c \simeq 10^{10} \cm$ in  resistive MHD model.  
 The evolutional tracks of magnetic energy and angular velocity begin to depart from each other at $n\simeq 10^{12}\cm$.

 The magnetic energy in resistive MHD model begins to decrease at $n_c \simeq 10^{12} \cm$, and becomes about $10^{-4}$ times smaller than thermal energy in resistive MHD model at $n_c \simeq 10^{17}$ because magnetic field lines are extracted from the high density region by the Ohmic dissipation.
 Conversely, the angular velocity in resistive MHD model becomes 10 times larger than that in  ideal MHD model at $n_c \simeq 10^{20} \cm$.
 Since the magnetic braking in  resistive MHD model is less effective than that in  ideal MHD model for weak magnetic field, the angular momentum in  resistive MHD model is  removed slightly compared with  ideal MHD model.
After central density reaches $n_c \simeq 10^{16}\cm$, the equation of state becomes soft (see equation~[\ref{eq:eos}]) reflecting the dissociation of hydrogen molecules at $T \simeq 2\times 10^3$ K, and collapses rapidly, i.e., the second collapse begins. 
By this epoch, the central temperature becomes so high that the thermal ionization of Alkali metals reduces the resistivity and so that Ohmic dissipation becomes ineffective.
 Thus, the magnetic field becomes strong again as central region collapses.
 The second core \citep[cf.][]{larson69} formed at $n_c \simeq 10^{21} \cm$ both in  resistive and ideal MHD models.
 The magnetic field strength increases rapidly after the second core formation epoch ($n\gtrsim 10^{21} \cm$), because the shearing motion between the second core and ambient medium amplifies the toroidal magnetic field around the second core.

\begin{figure}
\plotone{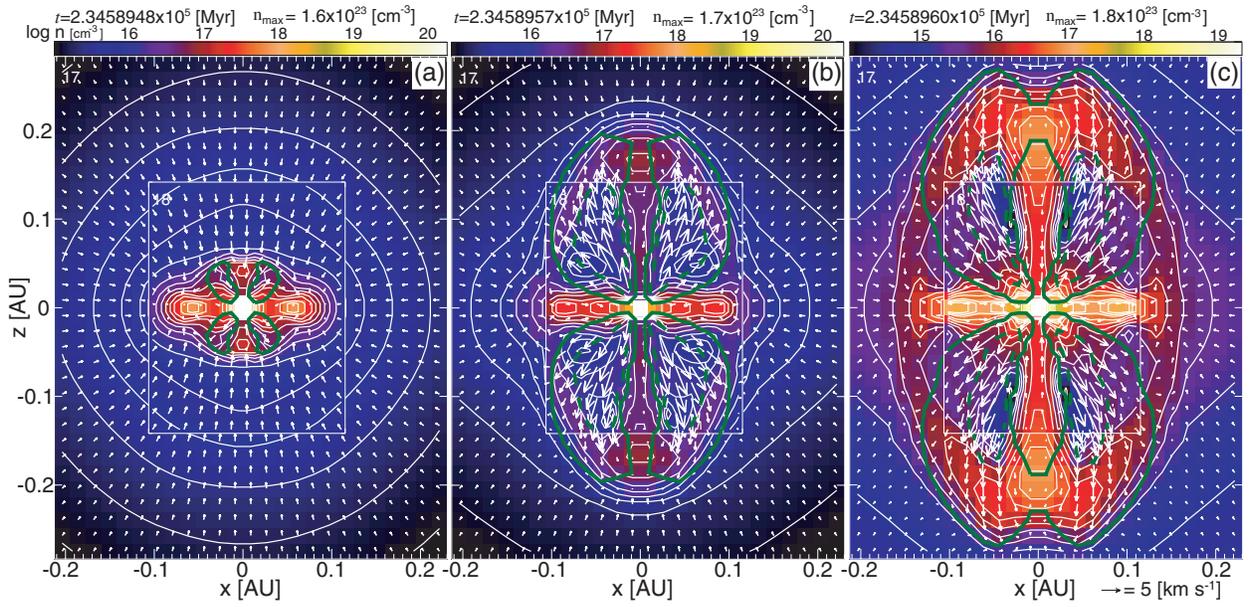}
\caption{
The density (color-scale and white contours) and velocity distribution (arrows) on the cross-section in the $y=0$ plane.
Panels (a)-(c) are snapshots at different stage, but the same grid level ($l = 17$ and $18$).
Green thick lines are velocity contours of outflow ($v_{out} > 3$ km s$^{-1}$; thick line; $v_{out} > $20 km s$^{-1}$; dashed  line).
}
\label{fig:2}
\end{figure}

 Figure~\ref{fig:2} shows the time sequence of the cloud in resistive MHD model after the second core formation ($n_c \gtrsim 10^{21} \cm$).
 Panels in this figure show only grids of $l$=17 and 18 levels to resolve jet-like structure, while we have resolved the deep interior at the central region with the grids of $l=1-21$ levels.
 A thin disk is formed inside the first core when the central density reaches $n_c \simeq 1.2 \times 10^{21} \cm$, and then a spherical core is formed within the disk.
 The disk has 19.4 $R_\odot$ of radial scale and the spherical core has 5.5 $R_\odot$ of radius.
 The mass of disk and core, respectively, increase up to $1.6\times 10^{-3} \msun$ and $7.2\times 10^{-3} \msun$ at the end of the calculation.
 We stopped calculation at 10.86 days after the second core formation ($n_c=10^{21} \cm$), because the time step becomes extremely small ($\Delta t \simeq$ 25 second !) for the Alfv\'en speed being high in the region just outside of the disk.
 The green contour in each panel denotes the iso-velocity curves representing the jet region in which the gas is outflowing from the core.
 The gas is outflowing from the spherical core with wide opening angle ($\simeq 45\degr$)
 and 25 km s$^{-1}$ of the maximum speed in Figure~\ref{fig:2}{\it a}.
 The jet expands to the vertical direction and cocoon-like structure is formed as shown in Figure~\ref{fig:2}{\it b}.
 At the calculation end,  $7.2 \times 10^{-4} \msun$ of the gas is outflowing with 41 km s$^{-1}$ of the maximum speed.
The average accretion and outflow rate which are calculated after jet appeared are $3.4\times 10^{-3} \msun$ yr$^{-1}$, and $2.2 \times 10^{-5} \msun$ yr$^{-1}$, respectively. 
The bow shock structures are clearly seen at $z=\pm 0.24$ AU in Figure~\ref{fig:2}{\it c}.

 Figure~\ref{fig:3} shows the cloud structure and magnetic field lines in the bird's-eye view at the same epoch as Figure~\ref{fig:2}{\it c}.
 The structure of the jet is shown by blue iso-velocity surface.
 The jet is coiled by the magnetic field (stream lines).
 The red iso-surfaces above and below the second core inside blue iso-surface means the high-density region ($n_c>10^{16} \cm$) which indicates the strong mass ejection from the center that occurred in the past. 
 The ejected mass is also coiled by the magnetic field.

\section{Discussion}

 We start with a slowly rotating and strongly magnetized cloud in this paper.
 The ratio of rotational to gravitational energy is $4\times 10^{-3}$, while the magnetic energy is equivalent to the thermal energy in the initial cloud.
 The outflow driven by the first core is seen in \citet{tomisaka02}, \citet{matsu04}, and \citet{machida04,machida05b}, while these outflows does not appear in this model.
 This is because considerably slowly rotating cloud is assumed initially.
 However, we confirmed that the outflow is driven also by the first core in a more rapidly rotating cloud even if resistive MHD is adopted.

\begin{figure}
\plotone{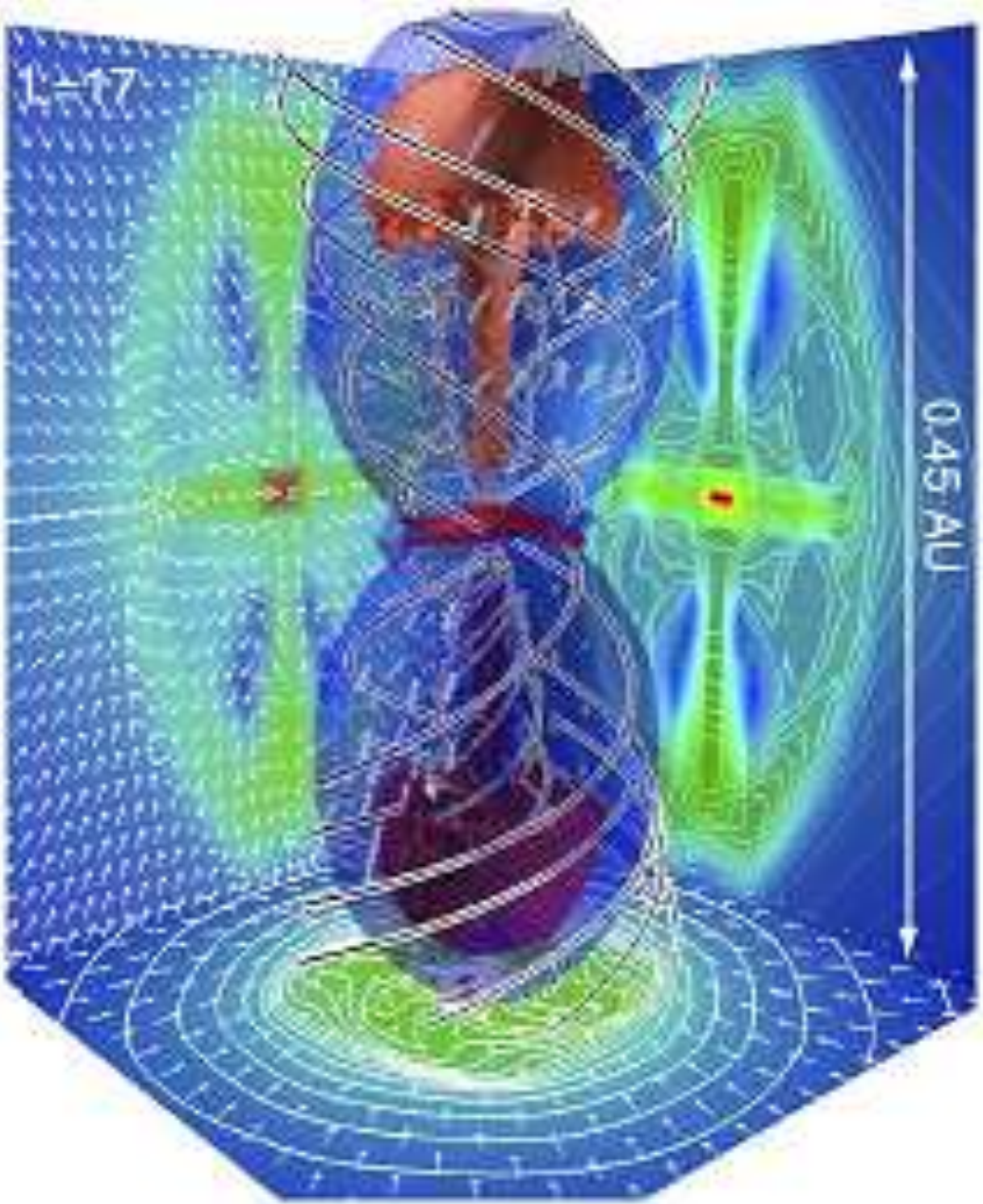}
\caption{
Bird's-eye view of $l=17$ grids in Fig.~\ref{fig:2} {\it c}.
The structure of high-density region ($n = 10^{16} \cm$; red iso-density surface), velocity vectors of outflow (thick arrows), and magnetic field lines are plotted.
The structure of the jet is shown by blue iso-velocity surface in which the gas is outflowing from the center.  
The density contours (false color and contour lines), velocity vectors (thin arrows) are projected in each wall surface.
}
\label{fig:3}
\end{figure}

The high-speed jet ($\sim 40\ $km\,s$^{-1}$) is driven by the second core both in resistive and ideal MHD model.
We have also calculated other models with various sets of parameters using both of resistive and ideal MHD methods.
The jet always appears in ideal MHD models, while jet does not appear in some resistive MHD models by $\sim 10$ days after the second core formation.
As shown in \S 3, the magnetic energy in  resistive MHD model is much smaller than the thermal energy when the second core is formed.
If the magnetic energy is larger or comparable to the thermal energy, jets are easily driven from the formed core by the magneto-centrifugally wind mechanism \citep{blandford82}.
However, this mechanism is inefficient when the magnetic energy around the protostar is extremely small.
In our resistive MHD model, the second core has a small magnetic energy.
Instead, the magnetic field in the formed disk around the central core is gradually amplified.
 We can see from the density structure in the $z=0$ plane projected into bottom wall of Figure~\ref{fig:3} that the spiral patterns are formed.
 When the central density exceeds $n_c > 10^{21} \cm$, non-axisymmetric (or spiral) patterns begin to grow inside the disk.
 It is considered that these patterns are caused by `bar model instability' \citep[e.g.,][]{durisen86} or magneto-rotational-instability \citep{balbus91}.
Because of these instabilities, the second core accretes not only
gas but also magnetic flux from the surrounding disk. 
This results in the increase of the magnetic pressure that is 
sufficient to launch the jet.
 The high speed jet in resistive MHD models seems to require these instabilities.
 Note that we had to stop the calculations in all models at $\sim 20$ days  after the second core formation due to the extremely small time step.
 Thus, it remains to determine whether or not and when the jet is driven ultimately in the models that have not shown it so far.
 However, jets may appear in models in which jet does not appear if we calculate further.
 The detailed analysis of the mechanism and  conditions for driving the jets will be investigated in our subsequent papers.

\acknowledgments
We have greatly benefited from discussion with Prof.~K. Tomisaka, Dr.~K. Saigo and Dr.~Y. Kato.We also thank T. Hanawa for contribution to the nested grid code.
Numerical calculations were carried out with a Fujitsu VPP5000 at the Astronomical Data Analysis Center, the National Astronomical Observatory of Japan.
This work was supported partially by the Grants-in-Aid from MEXT (15740118, 16077202, 16740115).

\clearpage

\end{document}